\documentclass[aps,a4paper, preprint, superscriptaddress,preprintnumbers,floatfix,nofootinbib,amsmath,amssymb]{revtex4}
\usepackage{cancel}
\usepackage{amsfonts}
\usepackage{amsmath}
\usepackage{amssymb}
\usepackage{color}
\usepackage{graphicx}
\usepackage{epsfig}
\usepackage{axodraw}
\usepackage{bm}
\usepackage{amstext,amssymb}
\usepackage{amsmath}
\usepackage{multirow}
\usepackage{graphicx}
\usepackage{dcolumn}
\usepackage[hyperfootnotes=false]{hyperref}
\usepackage{xspace}
\usepackage{braket}
\usepackage{color}
\usepackage{units}
\usepackage{comment}

\usepackage{slashed}
\usepackage[compat=1.0.0]{tikz-feynman}
\usepackage{soul}
\newcommand{\sym}{$U(1)_{B-L}$  }
\usepackage[normalem]{ulem}

\newcommand{\be}{\begin{equation}}
\newcommand{\ee}{\end{equation}}
\newcommand{\bea}{\begin{eqnarray}}
\newcommand{\eea}{\end{eqnarray}}

\usepackage[normalem]{ulem}
\newcommand{\lf}{\lambda_{eff}}
\newcommand{\Md}{M_{DM}}

\begin{document}

\title{Scalar Dark Matter and Radiative Dirac neutrino mass in an extended $U(1)_{B-L}$ model}
\author{Subhasmita Mishra,}
\email{subhasmita.mishra92@gmail.com}
\affiliation{School of Applied Sciences, Centurion University of Technology and Management, Odisha 761211}
\author{ Nimmala Narendra}
\email{nimmalanarendra@gmail.com}
\affiliation{Physical Research Laboratory, Ahmedabad, Gujarat 380009}
\author {Prafulla Kumar Panda}
\email{prafulla.k.panda@gmail.com}
\affiliation{Center of Excellence, High Energy and Condensed Matter Physics \\ Department of Physics, Utkal University, Bhubaneswar, India- 751004}
\author{Nirakar Sahoo}
\email{nirakar.pintu.sahoo@gmail.com}

\affiliation{Center of Excellence, High Energy and Condensed Matter Physics \\ Department of Physics, Utkal University, Bhubaneswar, India- 751004}

\preprint{}

\begin{abstract} 
 
We explore a gauged $U(1)_{B-L}$ extension of standard model with inclusion of three right-handed neutrinos of exotic $B-L$ charges to cancel the gauge anomaly. Non-trivial transformation of new particles under $B-L$ symmetry forbids the neutrino mass at tree level and hence a small Dirac mass can be generated radiatively at one loop with a doublet fermion and singlet scalar. We also discuss the phenomenology of a scalar dark matter, which can be obtained from the mixing of neutral CP even component of a doublet and real singlet scalar. An adhoc $Z_2$ symmetry is required in the current framework to stabilize the dark matter candidate. Presence of new particles with $Z_2$ odd charges and small mass splitting, makes the phenomenology more interesting by governing the relic density with co-annihilation processes. We explore the spin-independent direct detection constraints on dark matter via the scalar mediation. The new particle spectrum not only opens up new window for dark matter study but also satisfy the constraints from lepton flavor violating decay of $\mu \rightarrow e \gamma$.
\end{abstract}
\maketitle
\section{Introduction}

The Standard model (SM) of particle physics is the most successful theory to explain the interaction among the fundamental particles. But the enormous success of this model fails to explain certain experimental evidences like the small neutrino mass, existence of dark matter (DM) in the universe, etc. The discrepancies can be explained if one looks the physics beyond the SM framework. The existence of DM is confirmed by observations like the rotation curve of spiral galaxies, gravitational lensing \cite{Zwicky:1933gu,Rubin:1970zza} and are also responsible for the large scale structure formation of the universe. The relic abundance of the DM is precisely measured to be $\Omega_{DM} h^2 = 0.0223\pm 0.0002$ by satellite borne
experiments like WMAP \cite{wmap2013} and PLANCK \cite{Aghanim:2018eyx}. However, the particle nature like mass, spin, charge, etc., of DM candidate are still unknown. There exist several theoretical models to signify the particle components of DM \cite{Bertone:2004pz, Jungman:1995df}. Of them, weakly interacting massive particles (WIMP) are found to be one of the most promising candidate of DM. WIMP is assumed to exhibit weak interaction with the SM particles and hence was coupled to the thermal plasma during the early epoch of the universe. Later it decoupled from the thermal plasma once the interaction rate falls below the Hubble expansion rate and remains as a relic. There are many experiments running globally like LUX, XENON, PandaX, PICO, ATLAS, CMS, LZ \cite{Akerib:2013tjd,Aprile:2012nq,ATLAS:2019erb,PandaX-4T:2021bab,Griest:1990kh,Han:2019lux}to find the signal of such a DM candidate. But none of the experiments has been succeeded to find the signal of such a particles till date. The null observation of the above direct detection (DD) experiments puts a stringent constraint on DM parameter space.  The SM of particle physics is unable to explain the DM phenomenology.  One has to look physics beyond the SM for an explanation.

    Apart from the DM detection, a lot of information is missing in the neutrino sector. Such as the mass of neutrinos, the hierarchical ordering and also the Dirac or Majorana nature of the neutrinos, are remained as open questions for the physicists to explore \cite{Fukuda:2001nk, Bilenky:1980cx}. The smallness of neutrino mass can naturally be explained by the well known seesaw mechanism \cite{Minkowski:1977sc, GellMann:1980vs, Yanagida:1979as, Mohapatra:1979ia, Magg:1980ut, Lazarides:1980nt, Mohapatra:1980yp, Ma:1998dx, Konetschny:1977bn, Schechter:1980gr, Cheng:1980qt}, where the neutrinos develop a small Majorana mass. However, the Majorana nature of the neutrinos has not been established yet. The neutrinoless double beta decay experiments like GERDA, KamLAND-Zen, EXO, NEXT-HD etc \cite{Asaka:2018hyk,KamLAND-Zen:2016pfg,Giuliani:2019uno,GERDA:2019ivs,Alduino:2017ehq,Agostini:2019hzm} may probe the nature of neutrinos in future. Since the nature of neutrinos is still unknown, it is equally possible that these could be of Dirac type. The Dirac nature of the neutrinos is widely discussed in the literature \cite{Kanemura:2011jj,Kanemura:2011mw,Farzan:2012sa,Ma:2016mwh,Kanemura:2016ixx,Yao:2017vtm,Ma:2017kgb,Singirala:2017see,Borah:2018gjk,Calle:2018ovc,Yao:2018ekp,Ma:2019yfo,Dasgupta:2019rmf,Cai:2017jrq}. The general mechanism involved to explain a tiny Dirac neutrino mass is to forbid the tree level generation of neutrino mass either by a global $U(1)$ symmetry or by imposing some discrete symmetry. Thus a small Dirac mass can be obtained either by breaking the symmetry softly or the mass can be generated radiatively at loop level. It is interesting to focus on a model dependent discussion on combined study of neutrino and DMphenomenology. 
    
    In this work, we introduce three exotic right handed neutral fermions with ${\rm B-L}$ charges $5,-4,-4$ in order to cancel the triangle gauge anomalies. In addition to these, we have included two vector like fermion doublets ($N_{1,2}= (N^0_{1,2},N^-_{1,2})^T$), whose right and left handed components undergo same transformation under $SU(2)_L$ symmetry and hence does not contribute to triangle anomaly. While in the scalar sector, we have one doublet scalar ($\eta = (\eta^+ , \eta^0)^T$) and one complex singlet $\chi_2$ having $B-L$ charge $-3$ and $3$ respectively. Apart from these particles, a complete real singlet scalar $\chi_1$ is added as a DMcomponent. An extra $Z_2$ symmetry is required to stabilize the DM in the current framework. Under this symmetry only $\chi_1 , \eta$ and $N$  fields are odd and other new sector particles along with the SM fields are even. The neutrinos are massless at tree level due to the non-trivial $B-L$ charge of the right-handed neutrino fields (RHN). These acquire masses at one loop level after breaking of the symmetry of the theory. Once the $B-L$ symmetry is broken, the neutral CP even state of $\eta^0$ mixes with the real singlet scalar $\chi_1$ to form a singlet-doublet scalar DM candidate. Due to the mixing, the stringent constraint on singlet scalar DM \cite{Cline:2013gha, Athron:2017kgt} from the direct detection experiments can be evaded. The presence of new odd sector particles makes the DM phenomenology more interesting. The DM relic density is altered by these odd sector particles due to their contribution through the co-annihilation processes. These doublet scalar and fermion fields interaction rate falls within the weak interaction rate and hence the contribution is more if the mass difference between the DM and odd sector particles are kept small. We note that these particles were present in the early universe through their interaction with the SM particles in the thermal plasma. After decoupling from the thermal bath, their number density convert into the DM number density. In other words, we can actually get the relic density in the correct ball park even by the co-annihilation processes, when the annihilation processes are suppressed. These next to lightest stable particles (NLSP) does not contribute to the direct detection cross-section and opens new window for relic density.

    The article is structured as follows: Section~\ref{sec:mod} includes the details of the model and Lagrangian followed by the spontaneous symmetry breaking and mixing in the scalar sector. Radiative generation of neutrino masses is explained in the section~\ref{sec:neutrino}\,.  We commented on the constraints from lepton flavor violation in section~\ref{sec:lfv}\,.  In section~\ref{sec:dm}\,, we explored the DM phenomenology including the direct detection and the relic density and finally conclude in section~\ref{sec:conc}.

\section{The Model} \label{sec:mod}
We augmented the SM with a gauged \sym and $Z_2$ symmetry. The particle spectrum is extended with three right handed neutrinos $\nu_{R_i}$~($i=1,2,3$) with exotic $\rm B-L$ charges $5,-4,-4$ respectively. One complex singlet scalar $\chi_2$ having $B-L$ charge $3$ \,is introduced to break the $U(1)_{\rm B-L}$ symmetry at TeV scale and provides masses to the corresponding gauge boson. The real singlet $\chi_1$ and doublet $\eta$ scalars are included to form a suitable DMcandidate after mixing and stabilized by the $Z_2$ symmetry. Due to the exotic charges of right handed neutrinos, the tree level generation of neutrino mass is forbidden. In order to get the neutrino mass at one loop level, two vector like heavy Dirac fermion doublets $N_{1,2}$ are introduced. The particle content of the model and their respective charges under different gauge group is shown in the Table \ref{tab:prtcl} in detail. 
\begin{table}[h]
    \begin{tabular}{|c|c|c|c|c|}
      \hline
      Fields                     &  $SU(2)$    & $U(1)_Y$ & \sym & $Z_{2}$ \\
      \hline 
      $\ell_L \equiv (\nu_L, e_L)^T$         & 2  & -1   & -1   & +1 \\
      $e_R$                             & 1  & -2   & -1   & +1 \\
      $Q_L \equiv (u , d)_L^T$          & 2  & 1/3  & 1/3  & +1 \\
      $u_R$                             & 1  & 4/3  & 1/3  & +1\\
      $d_R$                             & 1  & -2/3 & 1/3  & +1\\
      $\nu_{R_1}$                       &1   & 0    & 5    & +1\\
      $\nu_{R_{2,3}}$                   &1   & 0    & -4   & +1 \\
      $N_{1} \equiv ( N^0_1, N^-_1)^T$  &2   & -1   & -1   & -1\\
      $N_{2} \equiv ( N^0_2, N^-_2)^T$  &2   & -1   & -1   & -1\\
      \hline
       $H$                              & 2  & 1    & 0    & +1 \\
      $\eta \equiv (\eta ^+ , \eta^0)$  & 2  & 1    & -3   & -1\\
      $\chi_1$                          & 1  & 0    &  0   & -1 \\
      $\chi_2$                          & 1  & 0    & 3    & +1\\
      \hline
    \end{tabular}
    \caption{Particle content of the model and their transformations under the gauge group $SU(2) \times U(1)_Y \times U(1)_{\rm B-L} \times Z_2$  . }
     \label{tab:prtcl}
  \end{table}

  
  Inclusion of these new particles also lead to new interaction terms in the Lagrangian, where the relevant terms in the fermion interaction Lagrangian is given by 
\begin{eqnarray}  \label{Lagrangian}
&&\mathcal{L}^{\rm fermion}_{\rm Kin.}=
      \overline{Q}_L i \gamma^\mu \left(\partial_\mu+ig \frac{\vec{\tau}}{2} \cdot \vec{W}_\mu 
      +  \frac{1}{3} i\,g^\prime \,B_\mu + \frac{1}{3} i\,g_\text{BL} \,Z_\mu^\prime\right) Q_L   \nonumber \\
&&+  \overline{u_{R}} i \gamma^\mu \left(\partial_\mu
      +   \frac{4}{3} i \,g^\prime \,B_\mu + \frac{1}{3} i\,g_\text{BL} \,Z_\mu^\prime \right) u_{R} \nonumber \\
&&+  \overline{d_{R}} i \gamma^\mu \left(\partial_\mu
      -  \frac{2}{3} i\,g^\prime \,B_\mu + \frac{1}{3} i\,g_\text{BL} \,Z_\mu^\prime \right) d_{R} \nonumber \\
&&+  \overline{\ell_{L}} i \gamma^\mu \left(\partial_\mu+i g \frac{\vec{\tau}}{2} \cdot \vec{W}_\mu 
      -   i\,g^\prime \,B_\mu - i\,g_\text{BL} \,Z_\mu^\prime \right) \ell_{L}            \nonumber \\
    &&+ \overline{e_{R}} i \gamma^\mu \left(\partial_\mu
      - 2 i\,g^\prime \,B_\mu -i\,g_\text{BL} \,Z_\mu^\prime  \right) e_{R}      \nonumber \\
&&+ \overline{\nu_{R_1}} i \gamma^\mu \left(\partial_\mu
       +5 i\,g_\text{BL} \,Z_\mu^\prime \right) \nu_{R_1} 
       + \overline{\nu_{R_2}} i \gamma^\mu \left(\partial_\mu
       -4 i\,g_\text{BL} \,Z_\mu^\prime \right) \nu_{R_2}\nonumber \\
&&+ \overline{\nu_{R_3}} i \gamma^\mu \left(\partial_\mu
       -4 i\,g_\text{BL} \,Z_\mu^\prime \right) \nu_{R_3}\;\nonumber\\
&&+ \overline{N_1} i \gamma^\mu \left(\partial_\mu
       - i\,g_\text{BL} \,Z_\mu^\prime \right) N_1\;+ \overline{N_2} i \gamma^\mu \left(\partial_\mu
       - i\,g_\text{BL} \,Z_\mu^\prime \right) N_2\;.
\end{eqnarray}

The Yukawa interaction Lagrangian for the new fields can be written as
 \begin{equation}
    - L_{Yuk}= \sum_{\alpha,\beta=1,2} M_{N_{\alpha \beta}} \overline{N_\alpha} N_{\beta} +\sum_{\alpha=2,3}\sum_{\beta=1,2} y^\prime_{\alpha \beta} \overline{N_\beta} \tilde{\eta}\nu_{R_\alpha}   + \sum_{\alpha=e,\mu,\tau}\sum_{\beta=1,2} y_{\alpha \beta}\overline{\ell_\alpha} \chi_1 N_\beta +{\rm H.c}. 
    \label{eq:yuk}
  \end{equation}
  with $\widetilde{H}=i\sigma_2 H^*$. The scalar sector is enriched with two doublets ($\eta$, $H$) and two singlets ($\chi_1$, $\chi_2$), where the full Lagrangian for these scalar fields is given as following
\begin{eqnarray}
\mathcal{L}^{\rm }_{\rm scalar} &=&
      \left(\mathcal{D}_\mu H \right)^\dagger \left(\mathcal{D}^\mu H\right) + \left(\mathcal{D}_\mu \eta \right)^\dagger \left(\mathcal{D}^\mu \eta \right)
      +\left(\mathcal{D}_\mu \chi_{1}\right)^\dagger \left(\mathcal{D}^\mu \chi_{1}\right)\nonumber \\
      &&
       +\left(\mathcal{D}_\mu \chi_{2}\right)^\dagger \left(\mathcal{D}^\mu \chi_{2}\right)
      +V\left(H,\eta,\chi_1,\chi_2 \right),
\end{eqnarray} 
here, the covariant derivatives can be written as
\begin{eqnarray} 
&&\mathcal{D}_\mu H = \partial_{\mu} H+i\,g \vec{W}_{\mu }\cdot \frac{\vec{\tau}}{2}\, H  \,+\, i\frac{g^{\prime}}{2}B_{\mu} H\, , \nonumber \\
&&\mathcal{D}_\mu \eta =\partial_{\mu} \eta+i\,g \vec{W}_{\mu }\cdot \frac{\vec{\tau}}{2}\, \eta \,+\, i\frac{g^{\prime}}{2}B_{\mu}\eta -3i g_{\rm BL} \,Z_\mu^\prime \eta , \nonumber \\
&&\mathcal{D}_\mu \chi_{1} =\partial_{\mu} \chi_{1}  \, , \nonumber \\
&&\mathcal{D}_\mu \chi_{2} =\partial_{\mu} \chi_{2} + 3 i g_{\rm BL} \,Z_\mu^\prime \chi_{2}\,,.
\end{eqnarray}
and $V(H,\eta , \chi_1 , \chi_2) $ being the scalar potential is given by
  \begin{eqnarray}
  V&=&-\mu^2_H H^\dagger H+\lambda_H  (H^\dagger H)^2 + \mu^2_\eta \eta ^\dagger \eta + \lambda_\eta  (\eta ^\dagger \eta)^2 + \mu^2_1 \chi_1^2 + \lambda_1  (\chi_1)^4 \nonumber  \\
&&+ \mu^2_2 \chi^\dagger_2 \chi_2 + \lambda_2  (\chi^\dagger_2 \chi_2)^2+ \lambda_{\rm H1} (H^\dagger H) \chi_1^2 + \lambda_{\rm H2} (H^\dagger H)({\chi^\dagger_2} \chi_2) \nonumber\\
&&+\lambda_{\rm H\eta} (H^\dagger H) (\eta^\dagger \eta)+\lambda^\prime_{H\eta} (H^\dagger \eta) (\eta^\dagger H)+\lambda_{12} \chi_1^2 ({\chi^\dagger}_2 \chi_2)\nonumber \\ 
&&+\lambda_{\rm 1\eta}\chi_1^2  ({\eta^\dagger} \eta_2)+\lambda_{\rm 2\eta}({\chi^\dagger}_2 \chi_2)({\eta^\dagger} \eta_2) + \lambda_D((H^\dagger \eta)(\chi_1\chi_2)+ \rm h.c)
\label{eq:potential}
  \end{eqnarray}

\subsection{Spontaneous Symmetry breaking and Higgs mixing}
Spontaneous symmetry breaking of $\rm SU(2)_L\times U(1)_Y \times U(1)_{B-L}$ to $\rm SU(2)_L\times U(1)_Y$ is realized by assigning non-zero vacuum expectation value (vev) to the scalar singlet $\chi_2$. Later, the SM gauge group is broken spontaneously to low energy theory by the SM Higgs doublet $H$.  We emphasis here that, $\mu_\eta^2 , \mu_1^2>0$ and hence $\eta$ and $\chi_1$ are not getting any vev. However, the neutral CP-even states of these two fields mix with each other after the spontaneous symmetry breaking, lightest of which serves as a suitable candidate of DM in this framework.  The scalar sector can be written in terms of CP-even and CP-odd components after the symmetry breaking as:
\begin{align}
&H^0 =\frac{1}{\sqrt{2} }(v+h+ i A^0)\,, \nonumber \\
&\eta^0 =\frac{1}{\sqrt{2}}(h_e + i A_e)\,, \nonumber \\
& \chi_2 = \frac{1}{\sqrt{2} } (v_2+h_2+ i A_2)\,, \hspace{0.5cm} \nonumber\\
 \rm and \,\, & \chi_1 = \chi_1\,,
\end{align}

where, $\langle H\rangle=(0, v/\sqrt2)^T$, $\langle \chi_2\rangle=v_2/\sqrt2$.
The minimization conditions of the scalar potential Eq.~\ref{eq:potential} correspond to 
\begin{eqnarray}
 &&\mu^2_{\rm H} = -\frac{1}{2}\left[2 \lambda_H v^2 + \lambda_{H2} v^2_2\right] ,\nonumber\\
 &&\mu^2_{2} = -\frac{1}{2}\left[2 \lambda_2 v^2_2 + \lambda_{H2} v^2\right]
 \end{eqnarray}

We note that the imaginary part of the fields $H^0$ and $\chi_2$ are absorbed by the $SU(2)$ and $U(1)_{B-L}$ gauge bosons respectively. As a result the gauge bosons become massive.  There will be a mixing between the CP even states of $H^0$ and $\chi_2$ after the symmetry breaking.  The mass matrix can be written as :
\begin{align}
	M_{E}^2
	=
	\begin{pmatrix}
		2\lambda_{\rm H} v^2 	 & \lambda_{\rm H2}vv_2	\\
		\lambda_{\rm H2}vv_2	& 2\lambda_2 v^2_2	\\
	\end{pmatrix}. \label{h-phi1mix}
\end{align}
This $2\times 2$ matrix can be diagonalized by the usual rotation matrix as $U_E M^2_E U^\dagger_E= {\rm diag} \left[M^2_{H_1},M^2_{H_2}\right]$, where 
\begin{align}
	U_E
	=
	\begin{pmatrix}
		\cos{\beta}	 & \sin{\beta}	\\
		-\sin{\beta}& \cos{\beta}	\\
	\end{pmatrix}. 
\end{align}
The involved scalar couplings can be written as
\begin{eqnarray*}
&& \lambda_{\rm H}=\frac{1}{2 v^2}\left(\sin^2{\beta}{M^2_{\rm H1}}+\cos^2{\beta} M^2_{\rm H2}\right),\\
&& \lambda_{2}=\frac{1}{2 v^2_2}\left(\cos^2{\beta}M^2_{\rm H1}+\sin^2{\beta} M^2_{H2}\right),\\
&& \lambda_{\rm H2}=\frac{1}{vv_2}\left(\cos{\beta}\sin{\beta}(M^2_{\rm H2}-M^2_{\rm H1})\right).
\end{eqnarray*}
We assume here that $M_{H_1} = 125$ GeV and corresponds to the Higgs of the SM and $M_{H_2}$ is in TeV scale.  We also assume that the mixing between them is too small $<< \mathcal{O} (10^{-3})$ hence does not affect the DM phenomenology.\\ 

Similarly the mass matrix for CP even state of $\eta^0$ and $\chi_1$ can be written as:
\begin{equation}
  \begin{pmatrix}
    \chi_1  &  \eta^0
  \end{pmatrix}
  \begin{pmatrix}
    M_{\chi_1}^2        & \lambda_D v v_2 \\
    \lambda_D v v_2   & M_{\eta^0}^2
  \end{pmatrix}
  \begin{pmatrix}
    \chi_1 \\
    \eta^0
  \end{pmatrix}
\end{equation}
To diagonalising the above mass matrix we use the rotation as:

\begin{equation}
\begin{pmatrix}
\chi_1\\
\eta_0
\end{pmatrix}=\begin{pmatrix}
\cos{\alpha} & -\sin{\alpha}\\
\sin{\alpha} & \cos{\alpha}
\end{pmatrix} \begin{pmatrix}
\zeta_1\\
\zeta_2 \,,
\end{pmatrix}
\end{equation}
with $\alpha$ being the mixing angle and it is given by:
\begin{equation}
  \tan 2\alpha = \frac{ 2 \lambda_D \, v \, v_2 } {M_{\eta_0}^2 - M_{\chi_1}^2}
  \label{eq:Dm_theta}
\end{equation}
The mass eigenstates can be written as
\begin{eqnarray}
&& \zeta_1=\chi_1 \cos{\alpha}  +\eta_0  \sin{\alpha}\nonumber\\
&& \zeta_2=-\chi_1 \sin{\alpha}  + \eta_0 \cos{\alpha} \,,
\end{eqnarray}
with masses $M_{\zeta_1}$ and $M_{\zeta_2}$ respectively.
In our analysis of the DM phenomenology, we will use $\alpha$ as a free parameter along with the masses of the odd sector particles. Hence $\lambda_D$ will be no more independent and using Eq.~\ref{eq:Dm_theta}, it can be expressed as :

\begin{equation}
  \lambda_D = \frac{(M_{\zeta_2}^2 - M_{\zeta_1}^2) \, \sin 2\alpha } {2 v v_2}
  \label{eq:ld}
\end{equation}
We consider $\zeta_1$ to be the lightest particle with mass $M_{\zeta_1} = \Md$ and serves as a DM candidate in our model. Then $\zeta_2$ field acts as the NLSP in our model.
\section{Comment on Radiative Neutrino Mass } \label{sec:neutrino} 
\begin{figure}[!ht]
    \includegraphics[height=35mm,width=50mm]{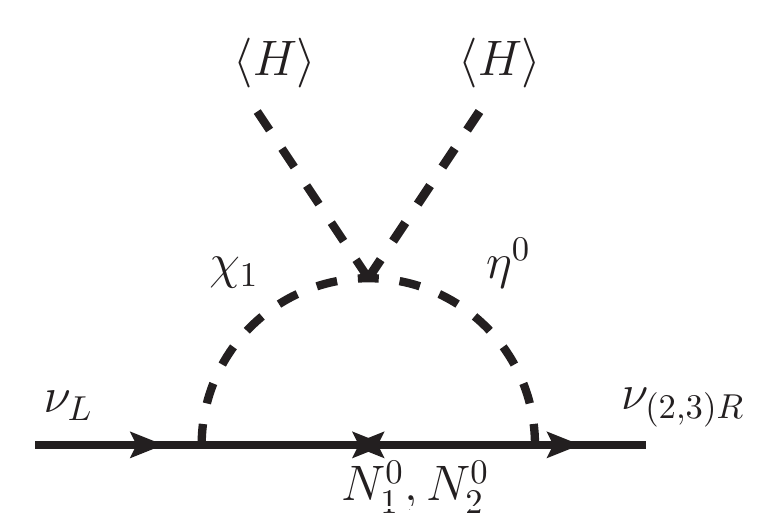}
    \caption{Radiatively generated neutrino masses.}
     \label{fig:neu_mass}
    \end{figure}

The right handed neutral fermions $\nu_{R_i}$ ($i$=1,2,3) have non trivial $B-L$ charges ($5,-4,-4$) and hence the tree level neutrino masses are forbidden by the symmetry. We can have the Dirac masses for two of the small neutrinos at one loop level Fig.~\ref{fig:neu_mass}.
 Radiatively generated neutrino masses, from the diagram in Fig. \ref{fig:neu_mass} can be estimated as \cite{Ma:2006km,Nanda:2017bmi}
\begin{equation}
(\mathcal{M}_{\nu})_{ij} = \frac{y_{ik}y^\prime_{jk} M_N \sin{\alpha} \cos{\alpha} }{16\sqrt{2} \pi^2} \left ( \frac{M^2_{\zeta_1}}{M^2_{\zeta_1}-M^2_N} \text{ln} \frac{M^2_{\zeta_1}}{M^2_N}-\frac{M^2_{\zeta_2}}{M^2_{\zeta_2}-M^2_N} \text{ln} \frac{M^2_{\zeta_2}}{M^2_N} \right)\,.
\label{numass1}
\end{equation}
Here, $M_{\zeta_1}$ ($M_{\zeta_2}$) is the mass eigenstate obtained from the mixing of the singlet scalar $\chi_1$ and CP even component of $\eta$. $M_N$ is the mass of the vector-like fermions $N_i$ in the loop.
The mass splitting between the CP odd and CP even component is given by
\begin{equation}
M^2_{\zeta_2}-M^2_{\zeta_1} = \lambda_D v v_2 \sin{\alpha} \cos{\alpha}\,.
\end{equation}
%
%
%
The interaction Yukawa matrices for left and right-handed neutrinos are given by
\begin{eqnarray}
y=\begin{pmatrix}
y_{11} && y_{12} && 0\\
y_{21} && y_{22} &&0\\
y_{31} && y_{32} && 0\\
\end{pmatrix}, \hspace{3mm}
y^\prime=\begin{pmatrix}
0 && 0 && 0\\
y^\prime_{21} && y^\prime_{22} && 0\\
y^\prime_{31} && y^\prime_{32} && 0\\
\end{pmatrix}. \label{Yuk_matrix}
\end{eqnarray}

We have generated the radiative Dirac neutrino masses for the active neutrinos, where one of the neutrino remains massless. The Dirac mass matrix can be constructed from the availed coupling, which could be diagonalized by bi-unitary transformation with two different unitary matrices.
For sample benchmark values with Yukawas of same order $y_{\alpha \beta} \approx \mathcal{O}(0.0005)$, $M_N (M_{N_1} \approx M_{N_2}) =1000 $ GeV, $\sin\alpha =0.01$ and taking the mass of $M_{\zeta_1} =800$ and $M_{\zeta_2} =810$ GeV, a small neutrino mass of $0.1$ eV can be obtained. We note the light neutrino mass $\mathcal{O} (0.1 $eV) can be obtained either by tuning the Yukawa couplings or the new particle masses \cite{deSalas:2020pgw,Gariazzo:2018pei,Esteban:2020cvm}. We will discuss later on the effect of small Yukawa couplings on the DMphenomenology in consistent with the neutrino mass.

\section{Comment on Lepton Flavor Violation} \label{sec:lfv}
Lepton flavor violating decay processes have received great attention in recent time and these decays are very rare to be observed experimentally \cite{Chekkal:2017eka,Dev:2017ftk,Bu:2008fx,Dutta:2018fge,Mihara:2013zna}. Many experiments are looking forward in this direction and some of them have provided a stringent upper limit on these decay modes. In this context, $\mu \rightarrow e\gamma$ looks to be more preferred process to be measured with less background from observational point of view. The current experimental limit on this decay is Br$(\mu\rightarrow e\gamma)<4.2\times 10^{-13}$ from MEG collaboration \cite{TheMEG:2016wtm}.
\begin{figure}[!ht]
\includegraphics[height=40mm,width=70mm]{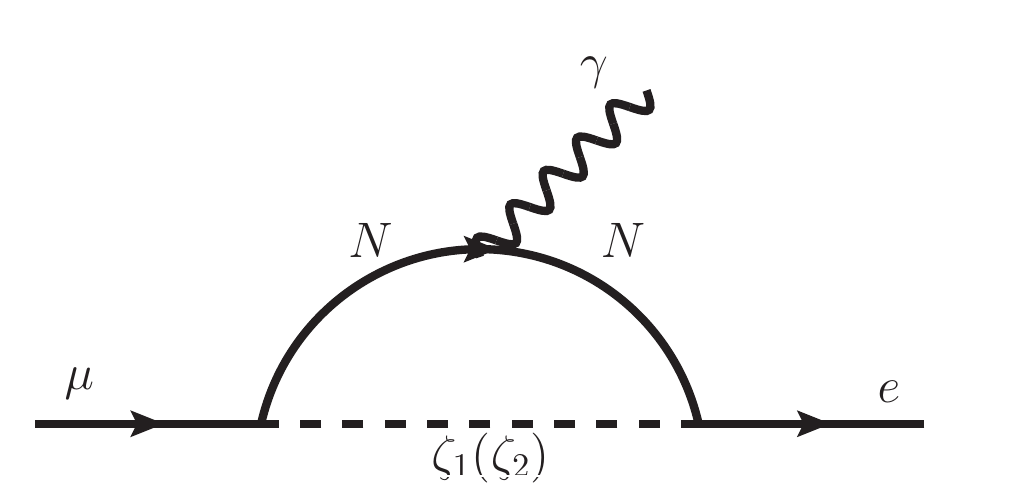}
\caption{Feynman diagram that violate lepton flavor.}
\label{fig:lfv_fey}
\end{figure}
The general expression for the decay width and branching fraction of process $(\ell_\alpha \rightarrow \ell_\beta \gamma)$ can be given as~\cite{PhysRevD.53.2442,PhysRevD.73.055003,De:2021crr}
\begin{eqnarray}
  && \Gamma(\ell_\alpha \rightarrow \ell_\beta \gamma) = \frac{\alpha_{em} m_\alpha^5 }{4} A_D^2 \\
 && Br(\ell_\alpha \rightarrow \ell_\beta \gamma)  =\tau_\alpha \times \Gamma(\ell_\alpha \rightarrow \ell_\beta \gamma)
\end{eqnarray}
where, $\tau_\alpha$ and $m_\alpha$ are the lifetime and mass of the decaying lepton respectively and $\alpha_{em}$ is the electromagnetic fine structure constant and $A_D$ is the dipole moment contribution. The expression for $A_D$ can be calculated from Fig.~\ref{fig:lfv_fey} as:
\begin{equation}
A_D=\sum_i \frac{y_{\alpha i} \,y^\star_{\beta i} \sin{\alpha} \cos{\alpha}}{2(4\pi)^2 M^2_{N}}\,\left(\frac{m_\alpha}{m_\beta} \right) f\left(\frac{M^2_N}{M^2_{\chi_1}}\right).
\end{equation} 
Here $y$ is the Yukawa coupling matrices, Eq.\,\eqref{Yuk_matrix}, $m_{\beta}$ is the final state lepton mass and $f(x)$ is the loop function, and is given by
\begin{equation}
f(x)=\frac{2+3x-6x^2+x^3+6x\,{\rm log}x}{6(1-x)^4}.
\end{equation}
\begin{figure}
  \includegraphics[height=50 mm, width =70mm]{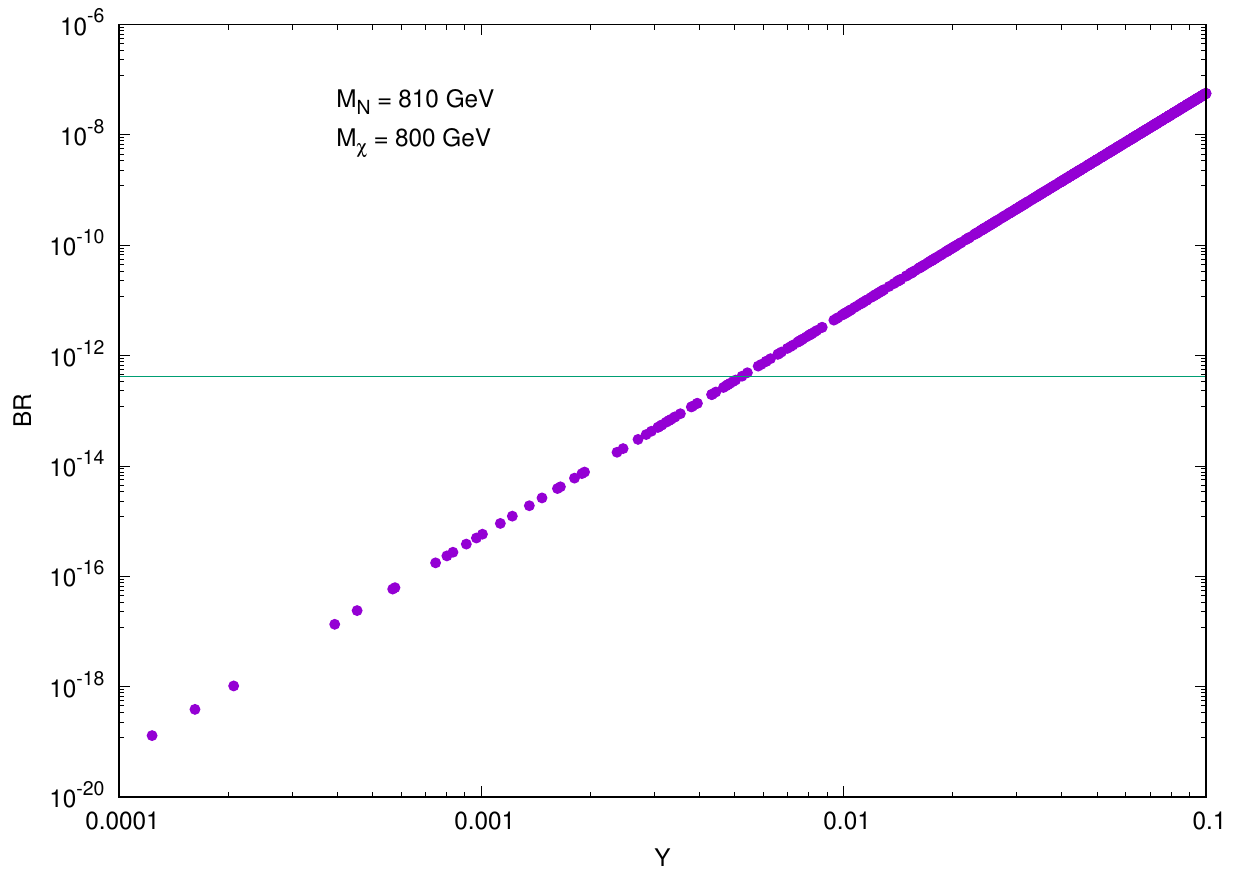}
  \includegraphics[height=50 mm, width =70mm]{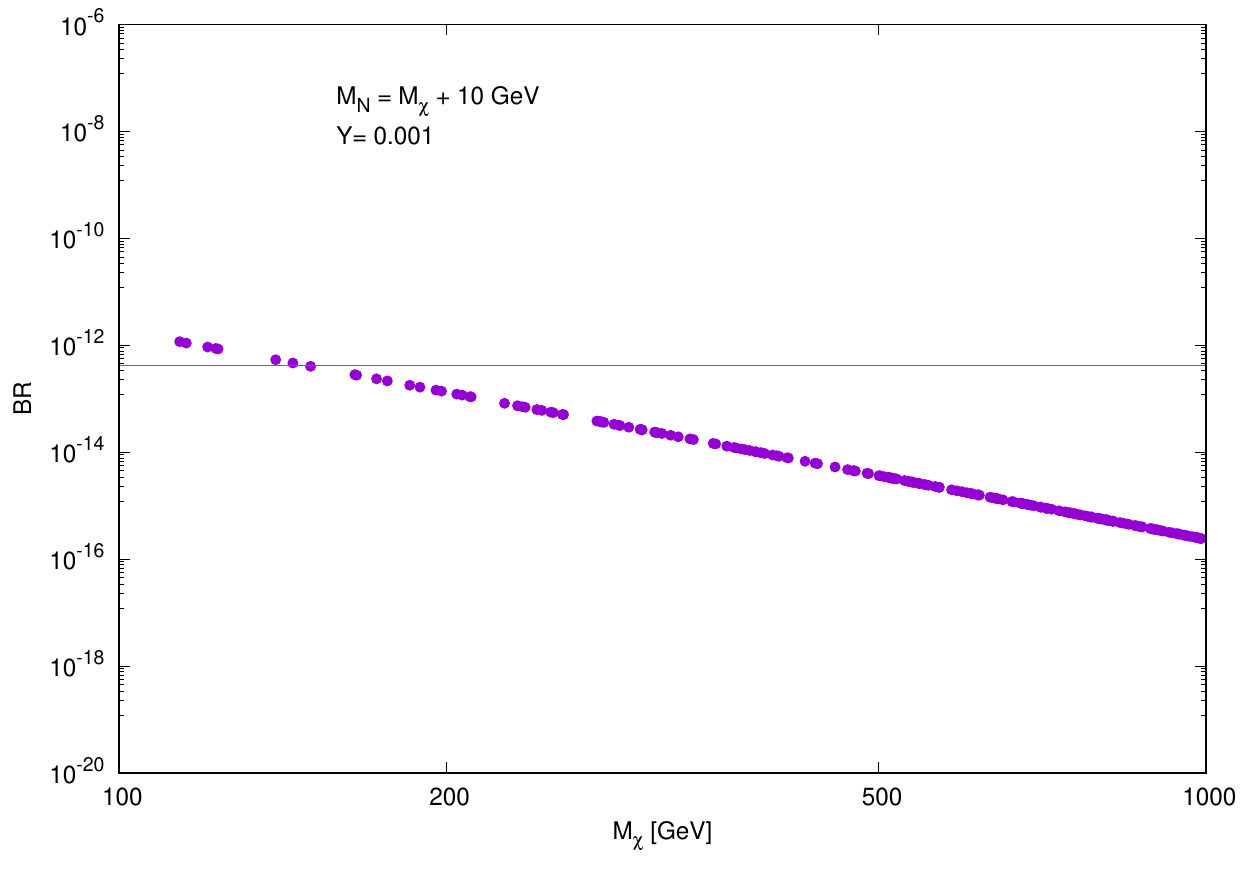}
  \caption{Branching fraction of $\mu \rightarrow e \gamma$ as a function of Yukawa coupling (Left) and Mass (Right).  The horizontal line corresponds to the current experimental bound.}
  \label{fig:lfv}
\end{figure}

We have plotted the branching fraction for $\mu \rightarrow e \gamma$ processes as a function of Yukawa coupling $Y$ (Left) and variation with mass $M_\chi$ (Right) in Fig.~\ref{fig:lfv}. The current experimental bound has been shown in the horizontal green line. For simplicity we have taken the similar order Yukawa couplings $y_{\mu e} \approx y_{ee} \approx Y$ through out our analysis. The VLF mass is fixed as $M_N =810$ GeV and the mass of $M_\chi =800$ GeV for the plot in Left, while $M_N = M_\chi +10 $ GeV is taken for the plot in right.  We can see that if we choose the mass of the scalar field as well as the VLF $M_N , M_\chi \geq 200$ GeV and $Y \leq 10^{-3}$, then the LFV constraint can be satisfied. However, this small Yukawa coupling will not contribute much to the muon $g-2$ value as clear from the Fig.~\ref{fig:lfv_fey} if one replaces $e$ field with $\mu$ field.  In other words, the stringent constraint on the Yukawa couplings comes from the LFV decays than the anomalous magnetic moment of muon.


\section{DM phenomenology } \label{sec:dm}
In the present framework, after the symmetry breaking the CP even state $\eta^0$ and the real singlet scalar $\chi_1$ mix with each other. The lightest particle among the new mass eigenstates ({\it i.e.,} $\zeta_{1}$ and $\zeta_{2}$) serves as a viable DM candidate, say $\zeta_1$. Since the DM is a mixture of singlet and doublet components, we will get the additional contribution to relic density by new annihilation and co-annihilation channels due to the presence of other odd sector particles. 
The Direct Detection (DD) can be possible via the $H_{1}$ and $H_{2}$ mediation. The required spin-independent (SI) elastic scattering cross-section can be achievable if we can tune the DM-DM-Higgs coupling.

\begin{figure}[!ht]
  \begin{tikzpicture}
\begin{feynman}[medium]
\vertex (a2);
\vertex [above left=of a2] (f1) {\(\zeta_1\)};
\vertex [above right=of a2] (f3) {\(\zeta_1\)};
\vertex [below = of a2] (c2);
\vertex [below left =of c2] (f2) {\({n,p}\)} ;
\vertex [below right =of c2] (f4) {\({n,p}\)};
\diagram* {
  (f1) -- [scalar] (a2),
  (f2) -- [fermion] (c2),
(a2) -- [scalar, edge label=\({H_1, H_2}\)] (c2),
(a2) -- [scalar] (f3),
  (c2) -- [fermion] (f4),
};
\end{feynman}
\end{tikzpicture}
  \caption{Feynman diagram of DD }
  \label{fig:fey_dd}
\end{figure}
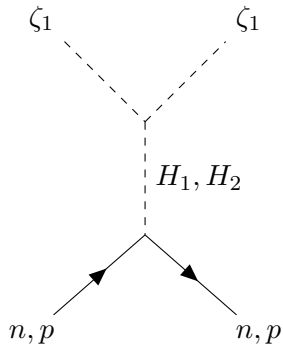

We note that the DM interacts with the quarks only through $H_1, H_2$ mediation as shown in Fig.~\ref{fig:fey_dd}.
By keeping the mixing between the SM Higgs and $\eta_0$ to be small, the contribution to DD cross-section can be suppressed for the diagram with $H_2$ mediation. We are then left with only a single channel, which contributes to the DD cross-section. We denote the coupling of $H_1 \zeta_1 \zeta_1$ vertex as $\lf$ and can be given as:
\begin{eqnarray}
  \lf &=& 2  \lambda_{H1} \cos^2\alpha \cos \beta \, v - 2\lambda _{12}  \cos^2\alpha \sin\beta \, v_2 + 2\lambda _{H\eta}    \cos \beta \sin\alpha^2 \,v  \\ \nonumber
  && + 2\lambda^\prime_{H\eta}  \cos \beta \sin^2\alpha\, v - 2\lambda _{\eta 2} \sin^2\alpha \sin\beta \, v_2 + \sqrt{2}   \lambda _D \cos \alpha \cos \beta \sin \alpha \, v_2 \\ \nonumber
  && - \sqrt{2}   \lambda _D  \cos \alpha \sin\beta \sin\alpha \, v.
 \end{eqnarray}
\begin{figure}[!h]
  \includegraphics[height=7 cm, width=7 cm]{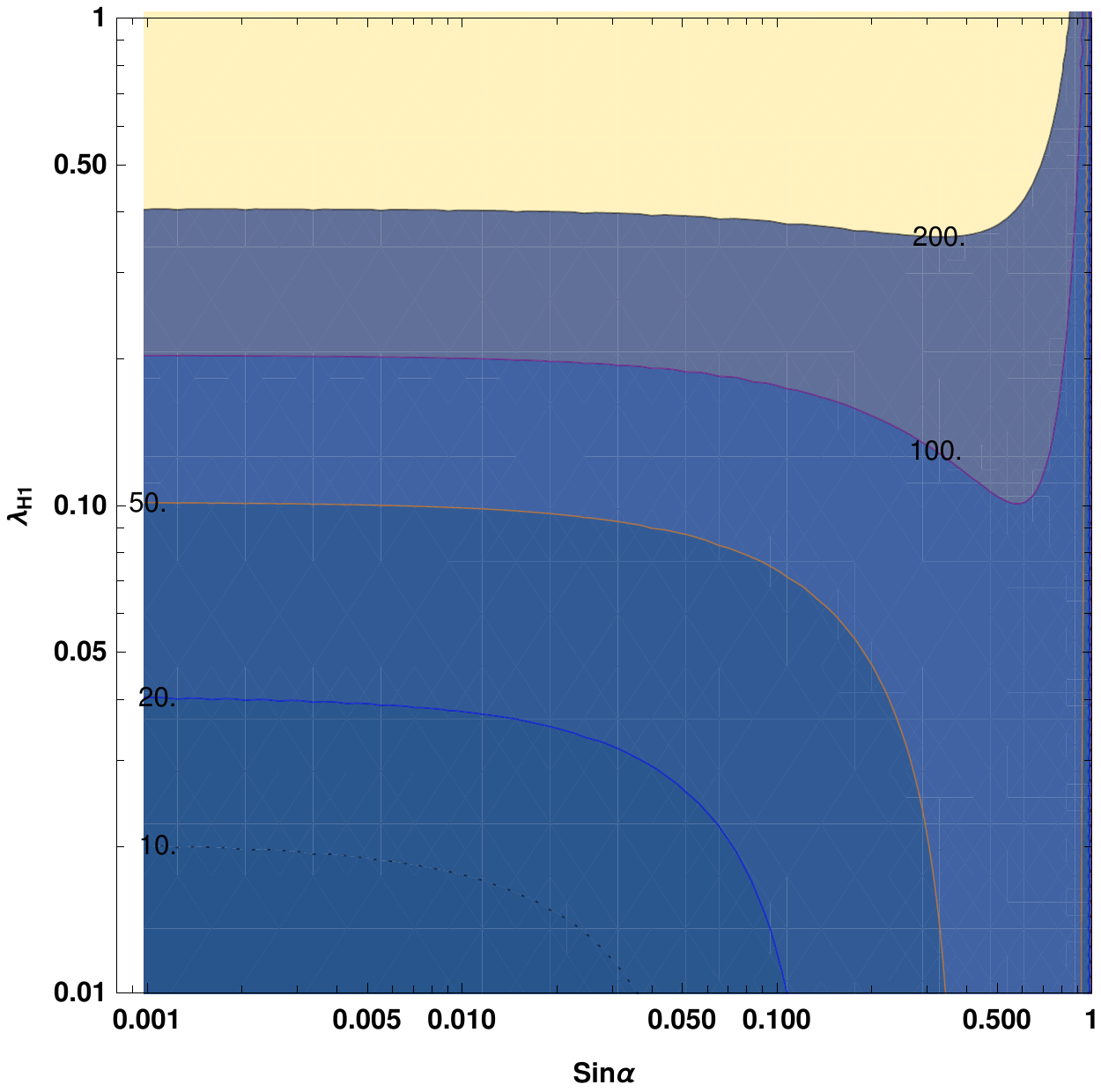}
  \caption{Contour plot showing the coupling of DM with the SM Higgs field ($\lambda_{eff}$) as a function of $\lambda_{H1}$ and $ \sin \alpha $.  We fix the parameter $\sin\beta=10^{-3}$ and all quartic couplings to be $\lambda_{12}=\lambda_{H\eta}=\lambda_{H\eta}'=\lambda_{\eta 2}=10^{-3}$, $\lambda_D=0.01$ and $v_2=10^4 $GeV.}
   \label{fig:leff}
  \end{figure}
The effective coupling $\lf$ depends on many parameters of the current model. However, we can see that the moment $\beta \rightarrow 0$, the effective coupling leads to the usual scenario of singlet-doublet scalar DM model \cite{Cohen:2011ec}.
To see the behavior, we have plotted different contours for $\lf$=10, 20, 50, 100, 200 GeV in the plane of $\lambda_{H1}$ and $\sin \alpha$ in Fig.~\ref{fig:leff}\,.
Since $v_2 >> v$, the terms containing $v_2$ contribute more to the $\lf$ irrespective of the small mixing angles $\alpha \rm \, and \, \beta$. But this can also be suppressed by taking the corresponding couplings to be very small ($\lambda_{12}, \lambda_{\eta 2} \rightarrow 0$).

The presence of the doublet fermions ($N_1, N_2$) open new window towards the contribution for DM relic density. Because of their $Z_2$ odd charge, the number densities of the these particles also control the number density of DM via co-annihilation.
This creates the model more interesting even in the absence of co-annihilation contribution from the scalar sector.  
For simplicity, we take certain assumptions for DM parameter space scan.
Here the DM mass $M_{DM}$ is a free parameter and we define the other odd sector particle masses with a mass splitting term $\Delta M = M_i-\Md$, where $M_i$ is the mass of the odd sector particles (NLSPs) $\eta$ and $N_{1,2}$.
 We assume the odd sector particles masses are all equal, unless otherwise explicitly specify. Note that the mass of the odd sector particles is dictated by $\Delta M$. The set of unknown parameters we consider to analyse the parameter space  are: $(\Md , \Delta M, \lf, \sin \alpha)\,$.

\subsection{Direct Detection }

The $Z_2$ stabilized real singlet scalar DM parameter space is severely constrained by DD experiments like LUX \cite{PhysRevLett.118.021303} and Xenon1T \cite{PhysRevLett.119.181301,Cline:2013gha,Athron:2017kgt}. As mentioned earlier, the DM interacts with the nucleus through the Higgs mediation. The relevant diagram contributing to direct detection is shown in Fig.~\ref{fig:fey_dd}\,. The SI cross-section is given by\cite{Jungman:1995df,Goodman:1984dc,Essig:2007az}
\be
\sigma_{SI}=\frac{4 \mu_R^2}{\pi} \left(Z f_p + (A-Z) f_n \right)^2 \,,
\ee

where, $\mu_R$ is the reduced mass of the DM-nucleon system, $Z,A$ are the atomic number and mass number of the nucleus respectively. The $f_p ,f_n$ are the effective coupling of the DM with proton and neutron of the target nucleus respectively. These effective couplings can be written as
\be
f_{p,n} = \sum_{q=u,d,s} f_{T_q}^{p,n} \alpha_q \frac{m_{p,n}}{m_q} \, + \, \frac{2}{27} f_{TG}^{p,n} \sum_{q=c,t,b} \alpha_q  \frac{m_{p,n}}{m_q} \, ,
\ee
with
\be
\alpha_q = \frac{ \lf} {M_{{H}_{1}}^2} \, \frac{m_q} {v}.
\ee

The different form factors are given by~\cite{Bertone:2004pz}
\begin{eqnarray}
&& f^p_{Tu}=0.020 \pm 0.004, \, f^p_{Td}=0.026\pm0.005, \,f^p_{Ts}=0.118 \pm 0.062\\ \nonumber
&& f^{(n)}_{Tu}=0.014\pm 0.004,~~ f^{(n)}_{Td}=0.036\pm 0.008, ~~f^{(n)}_{Ts}=0.118\pm 0.062.
\end{eqnarray}
The only unknown parameters in the DD cross-section are the effective coupling $ \lf $ and the mass of the DM, $M_{DM}$.  The DD experiments run all over the world in search of a DM signal, however no experiment has detected any signal till now. The simple reason behind this could be that the DM signal might be well below the sensitivity of the detectors. The null results from the DD experiments puts a constraint over the DMnucleus interaction cross-section and hence on the model parameters. Recently the stringent constraint on DD cross-section is obtained from LUX \cite{PhysRevLett.118.021303} and Xenon1T \cite{PhysRevLett.119.181301} experiments.
\begin{figure}[!h]
  \includegraphics[scale =0.7]{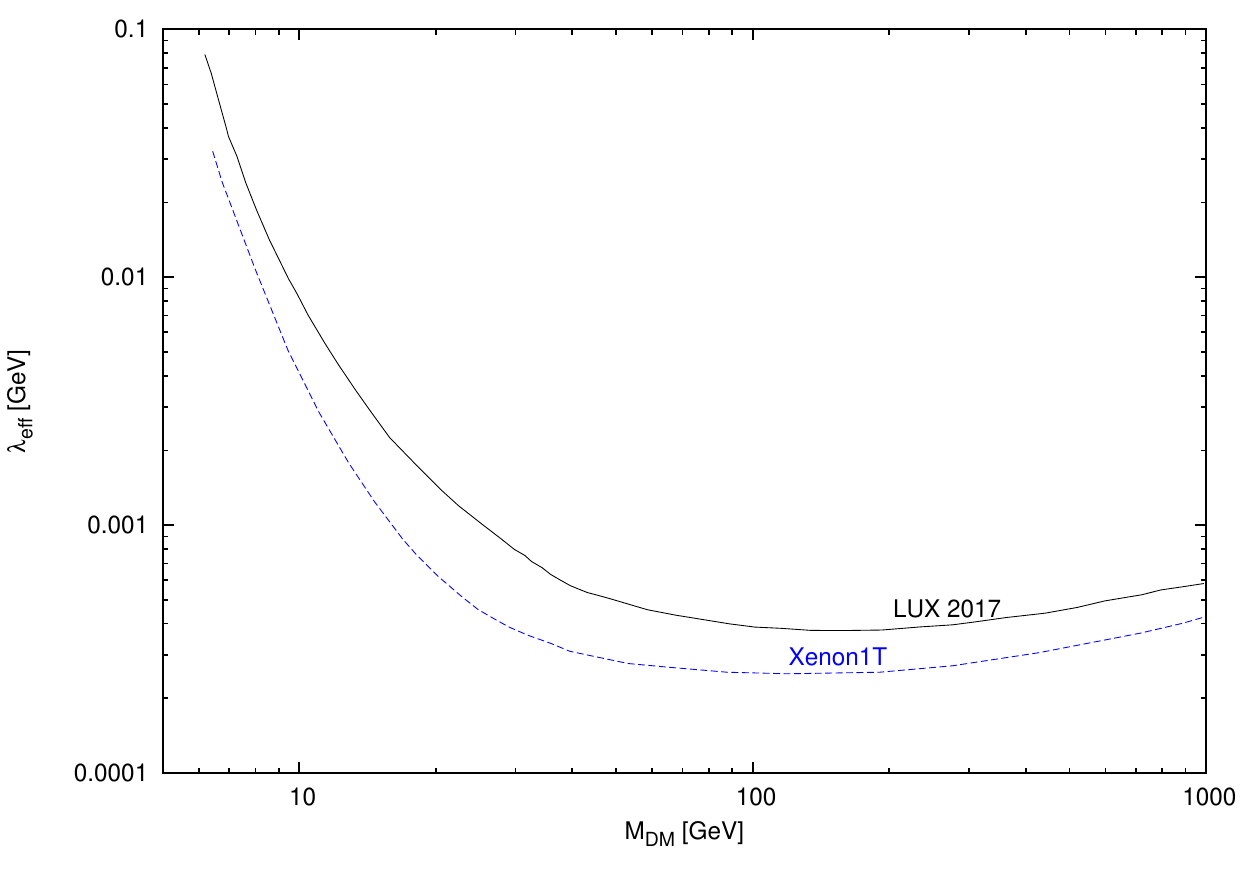}
  \caption{LUX (solid black) and Xenon1T (dashed blue) bound on the plane of $ \lf $ - $\Md$.  All points above the curves are ruled out.}
  \label{fig:DD_bound}
  \end{figure}

In Fig.~\ref{fig:DD_bound}\,, we have shown the DD experimental upper bound imposed on the model parameters, in the plane of $\lambda_{eff}$ vs $M_{DM}$, while considering the LUX (solid black) and Xenon1T (dashed blue) constraints. We can see that the region with $ \lf \gtrsim 3\times 10^{-4}$ is ruled out for DM mass $\Md > 40$ GeV. We can see in the next section that to obtain the correct relic density in the small mixing angle limit ($\beta \approx 10^{-3}$) through Higgs mediation, is currently ruled out by the DD experiments. However, the relic density will still be satisfied in this region, if co-annihilation contributions are considered. In other words, the co-annihilation channels dominate the relic density contribution keeping the DD constraint intact.


\subsection{Relic Density }
  The DM was present in the thermal bath in the early universe and decoupled, when its interaction rate falls below the expansion rate of the universe. After the decoupling, the DM remains as a relic and its density is only gets diluted by the expansion of the universe.  
The relic density of the DM directly depends on its interaction strength with the SM particles. There are also contribution to relic density from its co-annihilation processes, which involves the odd sector particles. The relic density of DM can be given as~\cite{Griest:1990kh}\,,
\begin{equation}
\Omega h^2 = \frac{1.07 \times 10^9 \, \rm GeV^{-1}}{J(x_F) \, g_\star ^{1/2} M_{Pl}} \, ,
\end{equation}

where 
\begin{equation}
J(x_F)= \int_{x_F}^\infty \frac{\langle \sigma |v| \rangle_{eff}}{x^2} dx \, ,
\end{equation}

with $\langle \sigma |v| \rangle_{eff}$ as the thermal averaged effective cross-section including both annihilation and co-annihilation processes. The effective cross-section expression can be given as 
\begin{equation}
\sigma_{eff} = \sum_{ij}^N \sigma_{ij} \frac{g_ig_j}{g^2_{eff}} \, (1+\Delta_i)^{3/2} \, (1+\Delta_j)^{3/2} {\rm exp} [-x (\Delta_i + \Delta_j)] \,,
\end{equation} 

where $i,j$ stands for the odd sector particles and $g_i,g_j$ are the spin degrees of freedom of $i,j$th particles respectively with
\begin{equation}
\Delta_i = \frac{M_i - \Md}{\Md} \, ,
\end{equation}
and 
\begin{equation}
g_{eff} = \sum_i^N g_i ( 1+ \Delta_i)^{3/2} exp [-x\Delta_i ]
\end{equation} 
is the effective spin degrees of freedom. For numerical calculation of relic density we used the package micrOMEGAs~\cite{Belanger:2014vza}. The model files required for micrOMEGAs is generated using LanHEP~\cite{Semenov:2014rea}.

To see the dependency of relic density on model parameters we divide our calculation into two parts (i) Relic density dominated by annihilation only, (ii) Relic density including co-annihilation contribution.


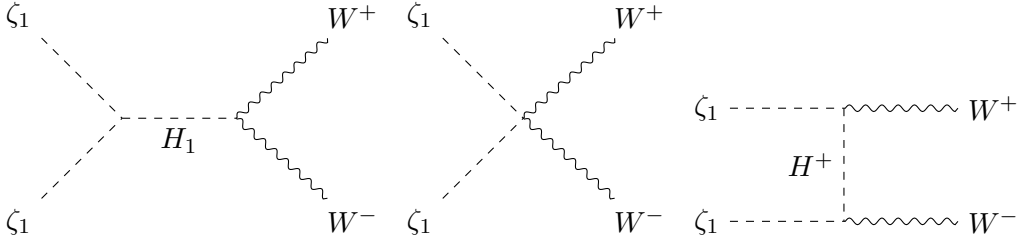
\begin{figure}
\raggedright
\begin{tikzpicture}
\begin{feynman}
\vertex (a);
\vertex [above left=of a] (f1) {\(\zeta_1\)};
\vertex [below left=of a] (f2) {\(\zeta_1\)};
\vertex [right = of a] (c);
\vertex [above right=of c] (f3) {\(W^{+}\)};
\vertex [below right=of c] (f4) {\(W^{-}\)};
\diagram* {
  (f1) -- [scalar] (a),
  (f2) -- [scalar] (a),
(a) -- [scalar, edge label'=\(H_{1}\)] (c),
(c) -- [boson] (f4),
(c) -- [boson] (f3),
};
\end{feynman}
\end{tikzpicture}
\begin{tikzpicture}
\begin{feynman}
\vertex  (x);
\vertex [above left=of x] (f10) {\(\zeta_1\)};
\vertex [below left=of x] (f20) {\(\zeta_1\)};
\vertex [above right=of x] (f30) {\(W^{+}\)};
\vertex [below right=of x] (f40) {\(W^{-}\)};
\diagram*{
(f10) -- [scalar] (x),
  (f20) -- [scalar] (x),
(x) -- [boson] (f40),
(x) -- [boson] (f30),
};
\end{feynman}
\end{tikzpicture}
\begin{tikzpicture}
\begin{feynman}[medium]
\vertex (a2);
\vertex [left=of a2] (f1) {\(\zeta_1\)};
\vertex [right=of a2] (f3) {\(W^{+}\)};
\vertex [below = of a2] (c2);
\vertex [left=of c2] (f2) {\(\zeta_1\)};

\vertex [right=of c2] (f4) {\(W^{-}\)};
\diagram* {
  (f1) -- [scalar] (a2),
  (f2) -- [scalar] (c2),
(a2) -- [scalar, edge label'=\(H^{+}\)] (c2),
(a2) -- [boson] (f3),
(c2) -- [boson] (f4),
};
\end{feynman}
\end{tikzpicture}
\caption{Dominant annihilation processes contributing to relic density.}
 \label{fey:ann}
\end{figure}
 \begin{figure}
   \includegraphics[ scale = 0.7] {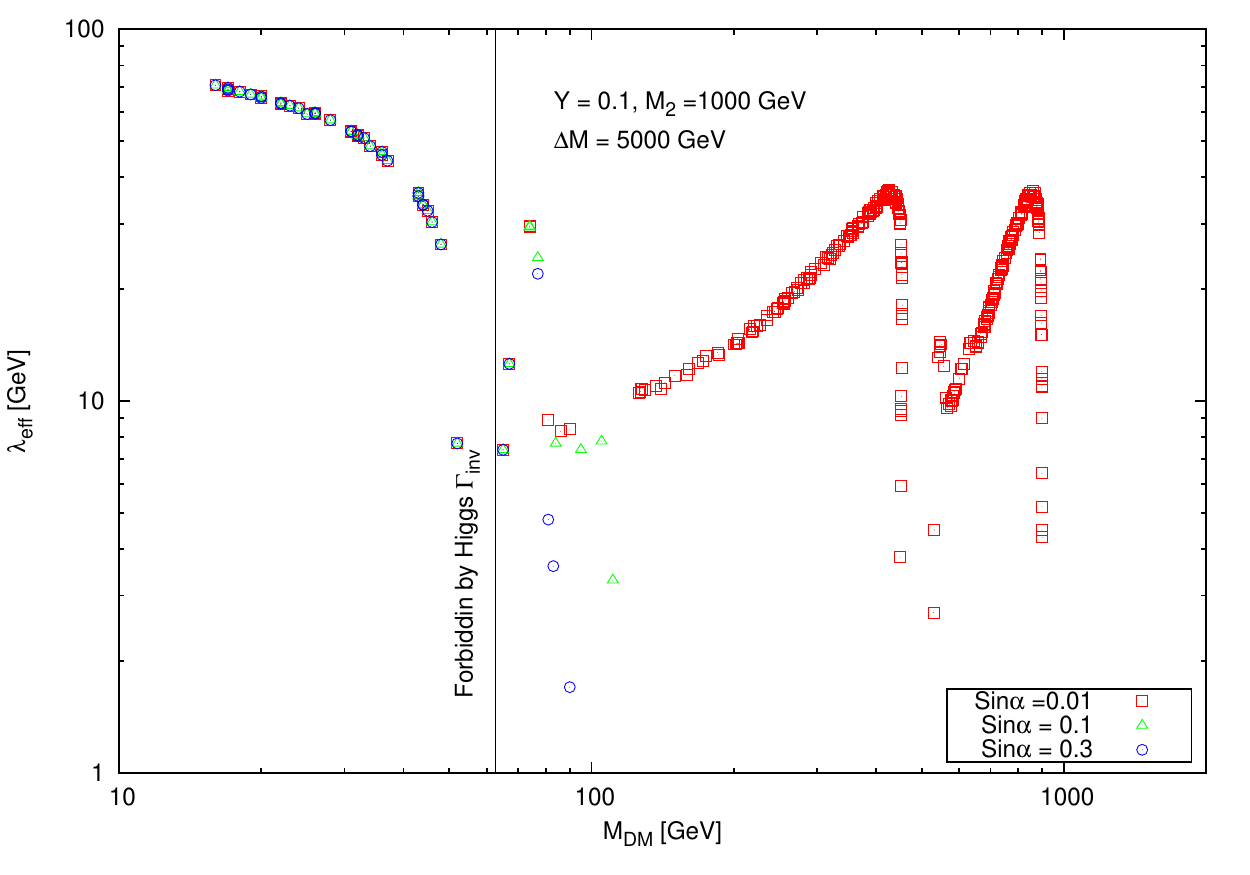}
   \caption{Required $\lf$ to satisfy the correct relic density for different $\Md$ for different $\sin\alpha = 0.01$(Red box), $\sin\alpha = 0.1 $(Green triangles) and $\sin \alpha =0.3$ (Blue circle).  The left to the solid vertical black line is disfavored by the Higgs invisible decay constraint. We have taken the Yukawa coupling $y=0.1$ and $M_2=1000$ GeV and the other remaining odd sector particle masses have taken 5 TeV  greater than the $\Md$.}
   \label{fig:ann_m_l}
 \end{figure}

 
 (i) {\bf{Annihilation}}: The co-annihilation processes are suppressed by introducing a large mass difference ($\Delta M= 5000$ GeV) between the DM and the odd sector particles by fixing $\sin\beta=0.001$. The relic density depends only on three parameters \{$\lf, \Md, \sin \alpha$\}. There are many processes contribute to the relic density. Among them the most dominant contribution comes from the channel $\zeta_1 \zeta_1 \rightarrow W^+ W^-$, when the DM mass is sufficient to produce this interaction. The dominant diagrams contributing to the relic density in this regime is shown in Fig.~\ref{fey:ann}\,.  We notice that the process with four point vertex (the second diagram of Fig.~\ref{fey:ann}\,) only depends on the $\sin\alpha$ apart from the SM gauge coupling and that dominates over the other subprocesses.  The $t$ channel process is a suppressed one.  As a result, for higher values of $\sin\alpha$ the relic density will be suppressed by the four point interaction subprocess. In order to show the dependency of relic density on model parameters, we have plotted the correct relic density points which satisfy the PLANCK data {\it i.e.,}
 \begin{equation}
   \Omega_{{\rm DM}} h^2 =0.120 \pm 0.001 \,,
 \end{equation}
 in the plane of $\Md$ and $\lf$ in Fig.~\ref{fig:ann_m_l}\, for different values of the mixing angle $\sin\alpha = 0.01$(Red box), $\sin\alpha = 0.1 $(Green triangles) and $\sin \alpha =0.3$ (Blue circle). We have taken the Yukawa coupling $y=0.1$ and $M_2=1000$ GeV. The solid vertical black line corresponds to the constraint from Higgs invisible decay width. The region left to this line is disfavored by the invisible Higgs decay width constraint \cite{Abe:2019wku,Baek:2014jga}. We see from the figure that there is a suppression near $\Md \sim 100$ GeV.  This region is contributed by the four-point process of Fig.~\ref{fey:ann}\, alone, hence the $\lf$ value is very small and this indicates the contribution through Higgs mediation is nullified. The other two suppressions corresponds to a resonance contribution from $H_2$ near $\Md$= 500 GeV and the other one around $M_{DM}=1000$ GeV corresponds to $\zeta_1 \zeta_1 \rightarrow H_2 H_2$ channel, which are independent of $\lf$. For $\sin\alpha =0.01$, a large region of parameter space gives the correct relic density. However for $\sin\alpha=0.1,\, 0.3$, it is very limited because of the four-point subprocess whose contribution depends on $\sin\alpha$, becomes dominant and hence a huge cross-section highly suppresses the relic abundance. But the contribution is less suppressed for small mixing angle $\sin\alpha\, \lesssim\, 0.01$, hence the relic density satisfy for a large range of $\Md$ and $\lf$. We emphasize here that most of the parameter space is disfavored by the direct detection constraint of DM (see Fig.~\ref{fig:DD_bound}\,). To overcome the problem we include the contribution of co-annihilation processes and see the effect on relic density in the following analysis.

 (ii) {\bf{Co-annihilation}:} Relic density of DM is also altered by the co-annihilation processes. In the present model, we have two doublet fermions $N_{1,2}$ and the doublet scalar $\eta$ being odd under $Z_2$ symmetry, which will contribute to the relic density through the co-annihilation channels.  Since the DM is the lightest stable particle (LSP), we denote all other odd sector particles as NLSPs. For simplicity, we assume that these NLSPs have equal mass and their mass is fixed by $\Delta M$.
 \begin{figure}[h!]
     \includegraphics[scale =0.6]{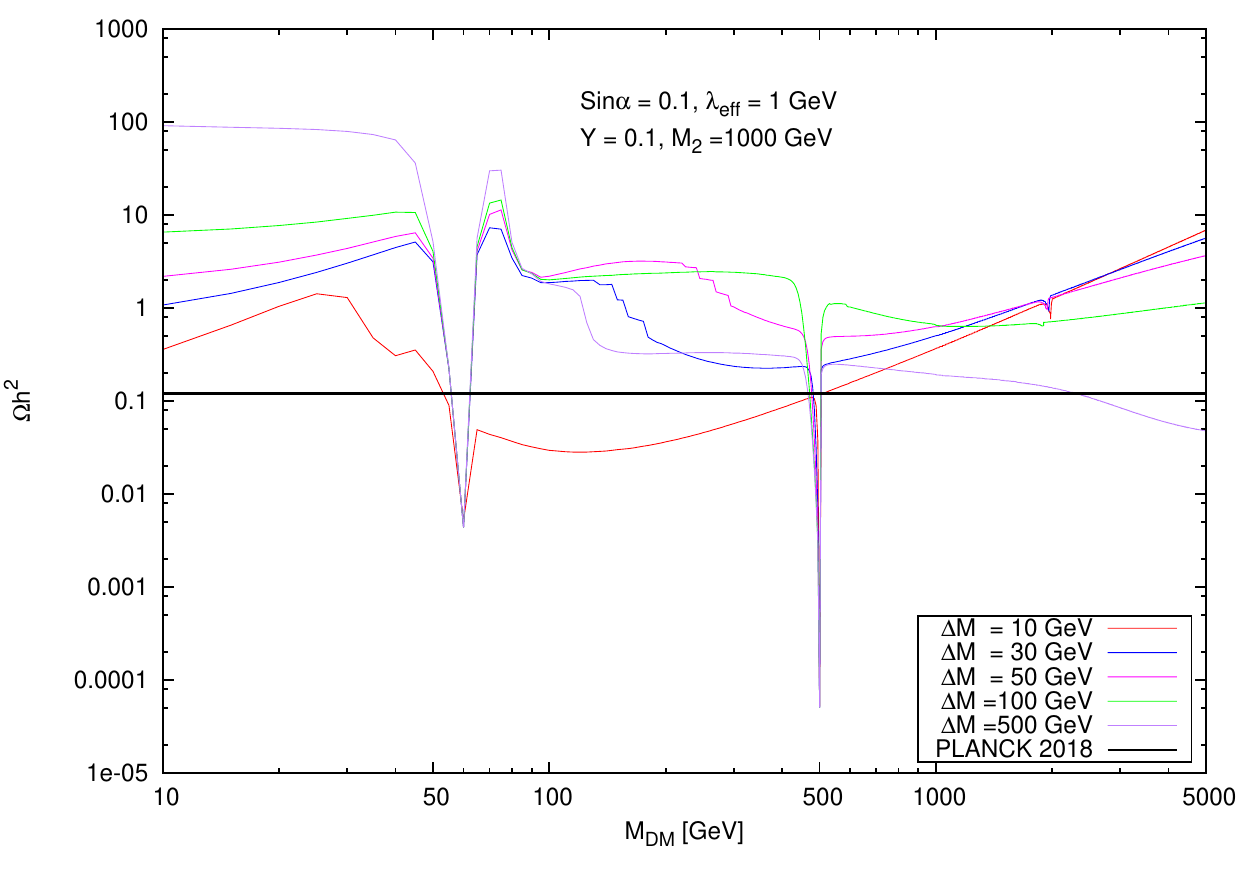}
     \includegraphics[scale =0.6]{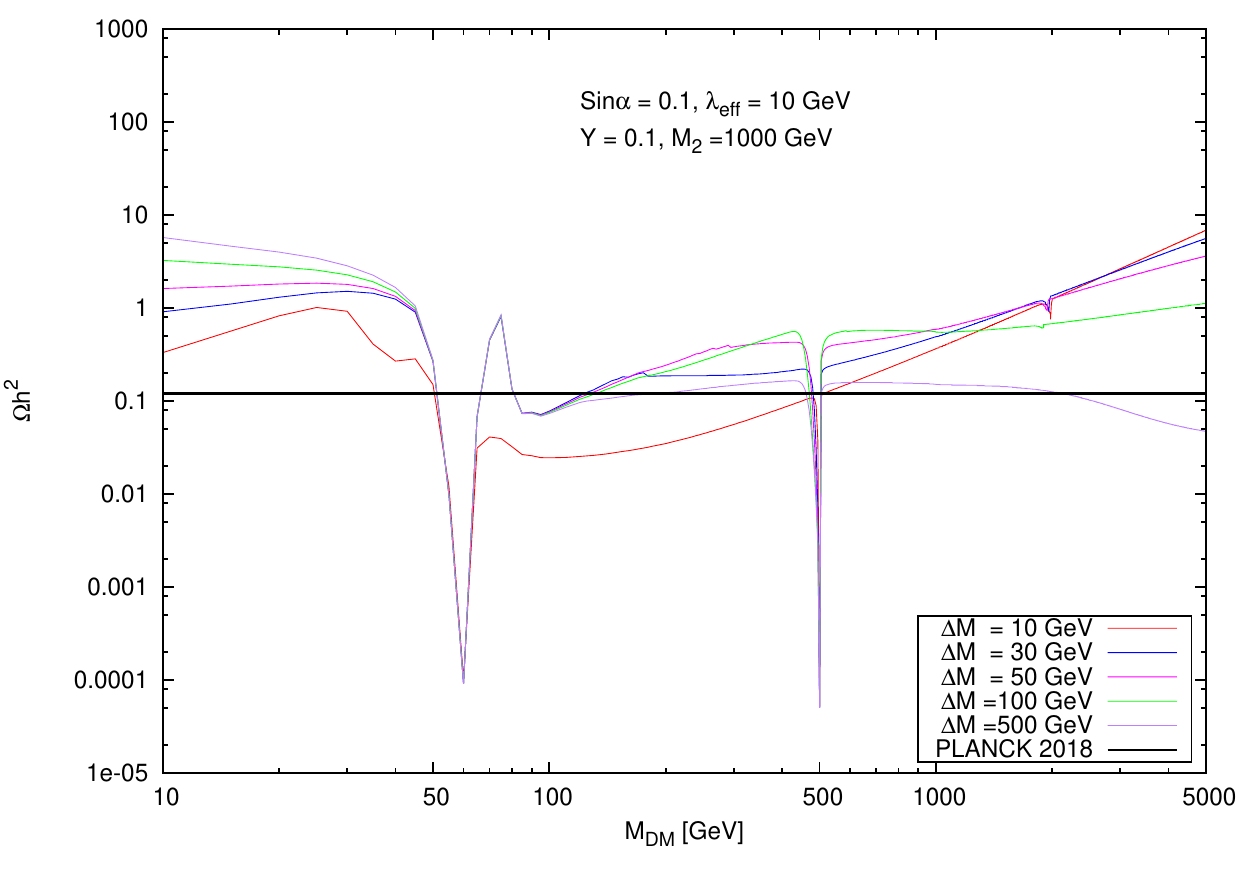}\\
     \includegraphics[scale =0.6]{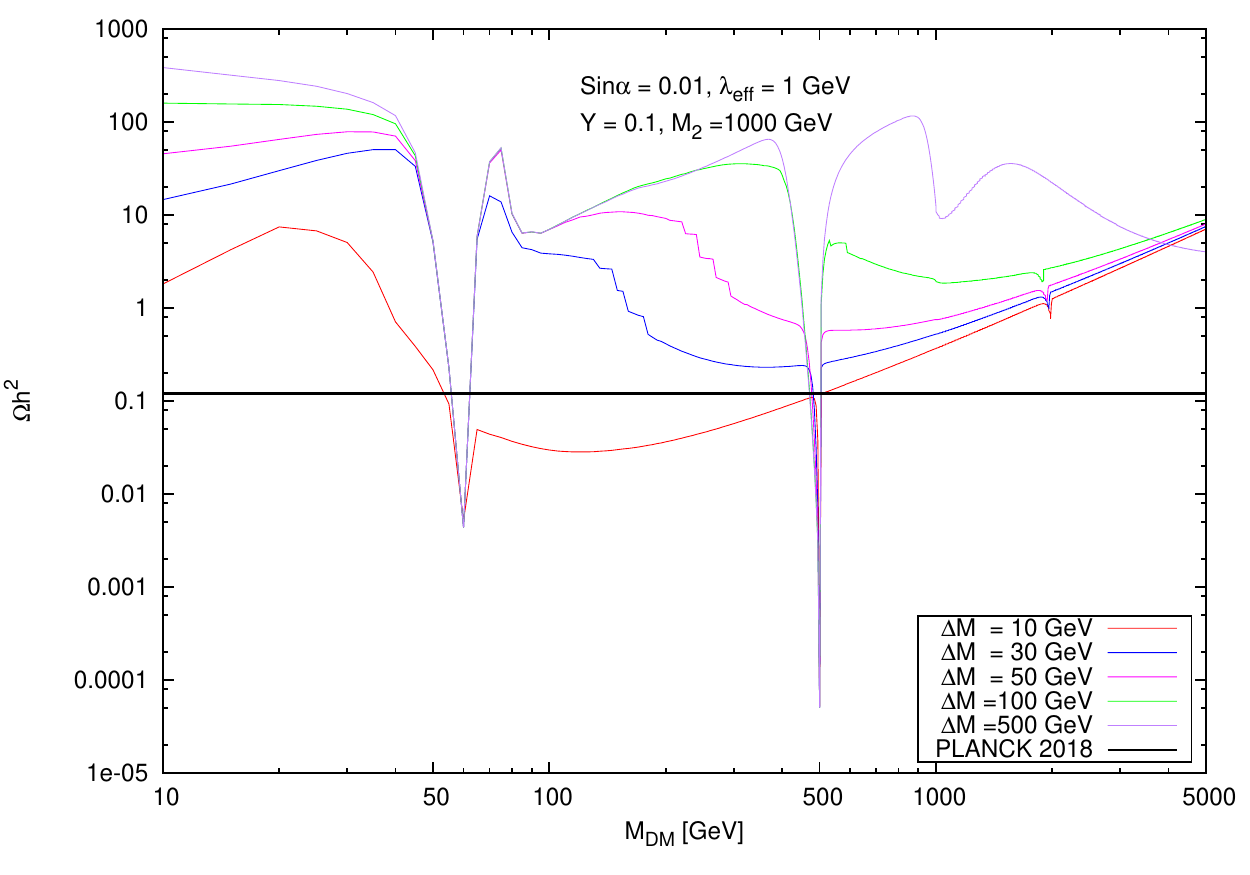}
     \includegraphics[scale =0.6]{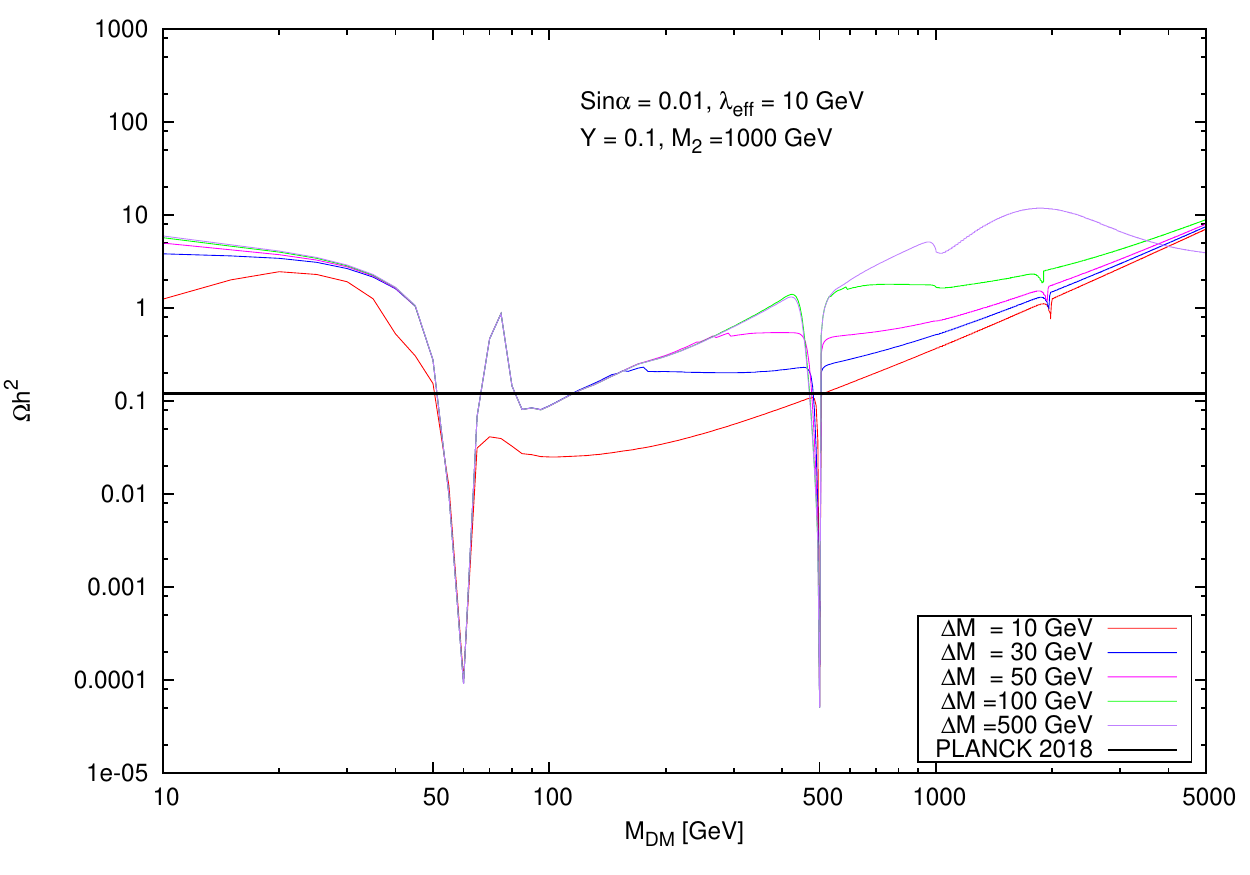}
   \caption{Relic density of DM as a function of $\Md$ for different values of  $\Delta M =10 $ GeV (Red), $\Delta M =30 $ GeV (blue), $\Delta M =50 $ GeV (magenta), $\Delta M =100 $ GeV (green), $\Delta M =500 $ GeV (purple).}
   \label{fig:con1}
   \end{figure}
To see the contribution of co-annihilation channels we have plotted the relic density of DM in the plane $\Md $ vs $\Delta M$ in Fig.~\ref{fig:con1}\,.
The upper two figures are plotted taking the mixing angle $ \sin \alpha =0.1 $ and the lower two figures are having $ \sin \alpha =0.01$. The figures in the first column correspond to $\lf =1 \,\rm GeV$ and the figures in the second column correspond to $\lf =10 \,\rm GeV$.

\begin{figure}
   \begin{tikzpicture}
  \begin{feynman}[medium]
\vertex (a);
\vertex [left=of a] (f1) {\(\zeta_1\)};
\vertex [right=of a] (f3) {\(H_1\)};
\vertex [below = of a] (c);
\vertex [left=of c] (f2) {\(\zeta_1\)};

\vertex [right=of c] (f4) {\(H_1\)};
\diagram* {
  (f1) -- [scalar] (a),
  (f2) -- [scalar] (c),
(a) -- [scalar, edge label'=\(\zeta_2\)] (c),
(a) -- [scalar] (f3),
(c) -- [scalar] (f4),
};
\end{feynman}
\end{tikzpicture}
   \caption{Dominant subprocess contributing to relic density}
    \label{fig:Feyh1h1}
 \end{figure}
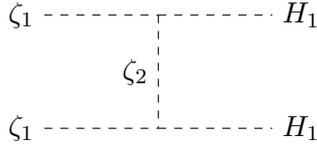

The relic density of DM is shown in different lines corresponds to different $\Delta M$ values $10 \,(\rm Red) ,30 \,(\rm Blue),50\,(\rm Magenta), 100 \,(\rm Green),500 \,(\rm Purple)$ GeV. The thick black solid horizontal line shows the correct relic density measured by PLANCK. We divide each figure for $\sin\alpha =0.1$ into three different regions \\
 Region I : $\Md<63 $ GeV,~ Region II : $63 \rm GeV <\Md< 500$ GeV and Region III : $500 \rm GeV <\Md<5000$ GeV. \\
 Region I is disfavored by Higgs invisible decay width and hence does not relevance in our case.
The co-annihilation processes are dominant for small values of $\Delta M < 50$ GeV. In this region, as we increase the mass splitting $\Delta M$, the co-annihilation contribution to the total cross-section decreases and hence relic density increases. But once the co-annihilation processes are suppressed for $\Delta M> 50$ GeV, only the annihilation channels contribute to the relic density as we discussed in the previous section. The cross-section of the process shown in Fig.\,\ref{fig:Feyh1h1} plays an important role with the $\Delta M$. The vertex $\zeta_1 \zeta_2 H_1$ depends on $\lambda_D$ which is directly proportional to $\Delta M$, see Eq.\,\ref{eq:ld}\,. Hence this cross-section increases with increase in $\Delta M$ and as a result the relic density decreases for $\Delta M =50-500$ GeV.  So the overall $\Omega h^2$ dependency can be seen as there is increment for $\Delta M = 10-50$ GeV and then decreases for $\Delta M = 50-500$ GeV. In region III, the above process also contribute to the relic density as like in region II upto $\Md = 2000$ GeV. The dip near $\Md = 500$ GeV corresponds to the contribution through $H_2$ mediated resonance processes.\\
 Again the mixing angle determines the nature of DM is more like a singlet or doublet. If the mixing angle is small then, it is more like singlet in nature hence there will be a suppression to cross-section. Similarly, if mixing angle is large, then it shows the doublet behavior and the cross-section will be enhanced. As a result, the relic density of DM is higher for lower $\sin \theta$ which can be read directly by comparing the top left figure with the bottom left one in Fig~\ref{fig:con1}.  We can also see that there is no such dependency of $\sin \alpha$ when $\lf =10$ GeV as the contribution of Fig.~\ref{fey:ann} is equal competent with other co-annihilation processes. This can also be ensured by comparing two figures having same $\sin \theta$ (0.01 or 0.1) with different $\lf =1$ GeV (left ones) and $\lf=10$ GeV (right ones) of Fig~\ref{fig:con1}. The larger the $\lf$ value larger contribution from the annihilation diagram hence smaller is the relic density.

\begin{figure}
     \centering
     \includegraphics[scale =0.6 ]{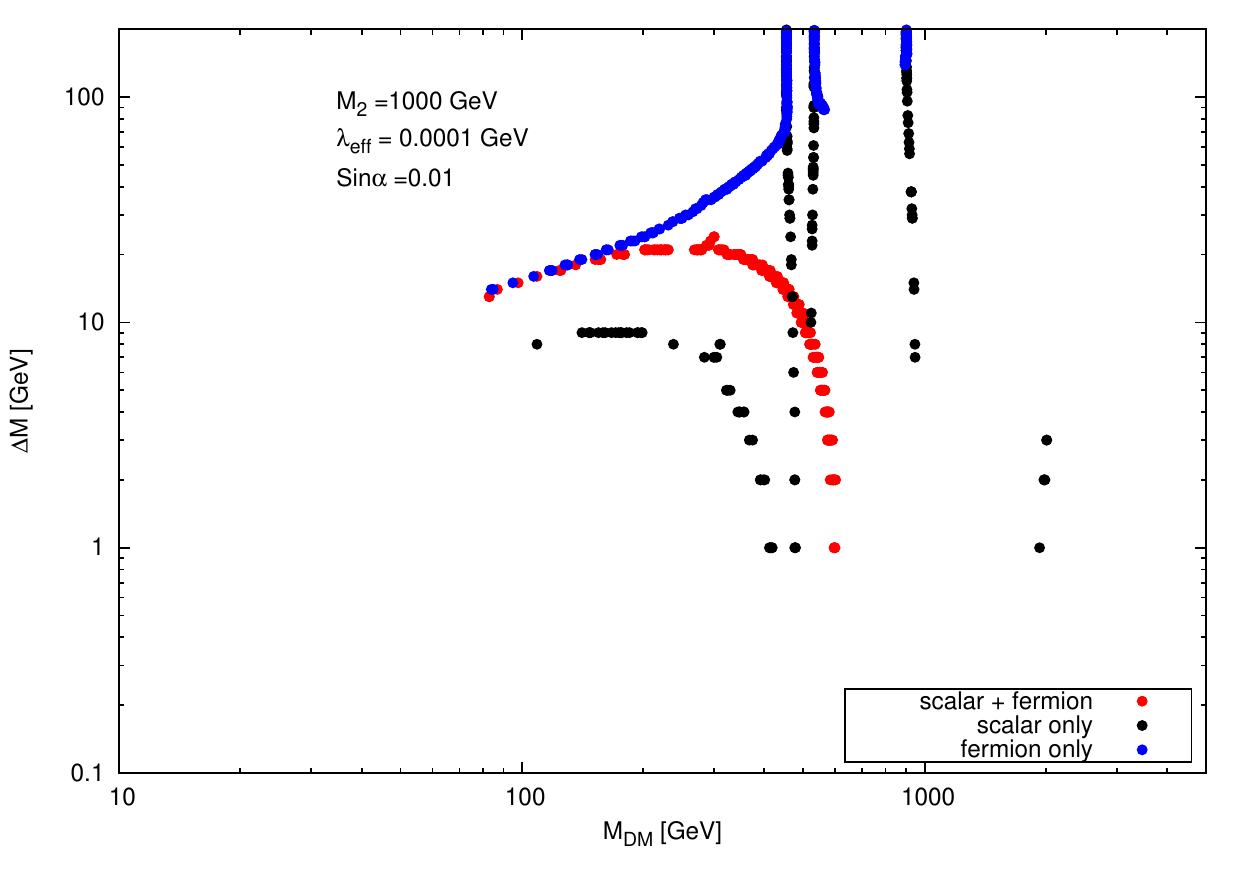}
     \includegraphics[scale =0.6 ]{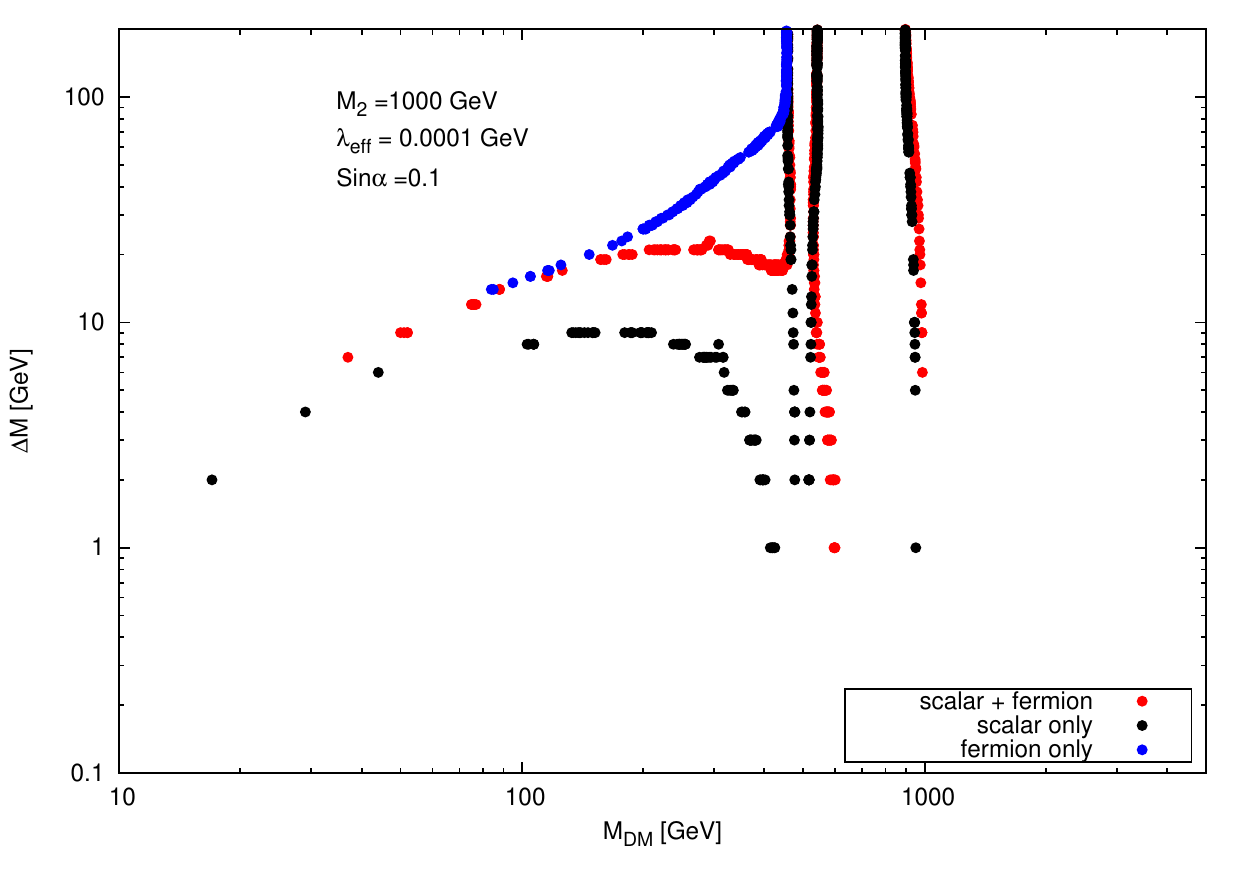}
     \caption{Allowed relic density in the plane of $\Delta M$ vs. $\Md$. }
     \label{fig:coan_del}
   \end{figure}

We point out here that both the scalar and the new fermion sector will contribute to the relic density. To see the effect of individual sector in the relic density calculation, we have plotted the relic density allowed as a function of $\Delta M$ vs $\Md$ in Fig.~\ref{fig:coan_del}. The left one corresponds to $\sin \alpha=0.01$ and the right one corresponds to $\sin \alpha =0.1$. In our calculation we assign a huge mass to one of the odd sector particles $ \Md / M_i  << 1 $, $i$ stands here for either scalar or fermion, hence their contribution to the relic density are nullified and the other sector mass can be fixed by $\Delta M$. The black dots
represent the allowed values of the parameters satisfying correct relic density constraint from PLANCK when the scalar particles are only involved in the co-annihilation processes. Similarly the blue dots correspond to the parameters allowed when fermion sector is only responsible for co-annihilation. Finally, we combined both the sectors and calculated the relic density for allowed parameter space which is shown in red dots in Fig.~\ref{fig:coan_del}. We again use the mass of $M_2=1000$ GeV and $\sin \alpha =0.01$ as before for this analysis. The value of $\lf$ is taken to be $10^{-4}$ GeV so that the constraint from direct detection experiments can be evaded. By comparing the two plots, we conclude that the independency of relic density on $\sin \alpha$ confirms that the relic density is mostly dominated by co-annihilation channels. We make sure that Yukawa coupling we considered is within the allowed value of LFV constraint as discussed in Fig.\,\ref{fig:lfv}\,.

\section{Conclusion} \label{sec:conc}
  In this present framework, we explored a gauged $B-L$ extension of the SM, where the particle spectrum includes three right-handed neutrinos having exotic $B-L$ charge $5,-4,-4$ to cancel the gauge anomaly. In addition to this, two doublet fermions $N_{1,2}$\,, a doublet scalar $\eta$, one complex singlet scalar $\chi_2$ and a real singlet scalar $\chi_1$ are added to the model in order to explain the neutrino mass and DM phenomenology simultaneously. An adhoc $Z_2$ symmetry is imposed under which, the SM fields are even and some of the new particles transform odd to contribute as DM candidates. The neutrinos are massless at the tree level due to the nontrivial $B-L$ charges of right handed neutrinos and hence radiative Dirac masses are obtained for two of the small neutrinos, while one of them still remains massless. We also constrained the model parameter from lepton flavor violating decay of $\mu \rightarrow e \gamma$. We found that the constraints on Yukawa coupling is not that stringent to satisfy the DM relic density. The experimental bounds from various neutrino and DMexperiments can be easily satisfied if the Yukawa coupling is made small $Y \leq 10^{-3}$ for TeV scale mass of the the vector like fermion and new scalar $\eta$. In our model, we consider the lightest mixed state out of the mixing of CP even component of doublet scalar $\eta$ and singlet scalar $\chi_1$ to be a suitable DMcandidate. Other odd particles such as the doublet fermions $N_{1,2}$ states are claimed to be the next to lightest stable particles and contribute to the relic density through co-annihilation processes. However, the real singlet DM parameter space is severely constrained from the direct detection experiments. The direct detection cross-section is made small to satisfy the bounds by taking the effective DM-DM-Higgs coupling $\lf$  to be small. We found that the DM relic density is satisfied by the Planck limit even with a small Yukawa coupling due to the dominant contribution from co-annihilation processes. The co-annihilation processes contribute dominantly to relic density when the mass splitting between the DM and co-annihilation partner is kept small $\Delta M \sim 100$ GeV.

   \section*{ Acknowledgement}
     NS would like to acknowledge RUSA 2.0, Ministry of Human Resource Management, India.

\bigskip

\bibliography{BL.bib}
\end{document}